\documentclass[twocolumn,prl,showpacs]{revtex4}

\usepackage{epsfig}
\usepackage{graphics}
\usepackage{graphicx}
\usepackage{dcolumn}
\usepackage{bm}

\def\LSCO{La$_{2-x}$Sr$_x$CuO$_4$ }
\def\LCO{La$_2$CuO$_4$ }
\def\cm-1{cm$^{-1}$}

\begin{document}

\title{Magnetic order in lightly doped \LSCO}
\author{A.~Gozar$^{1,2}$}
\email{gozar@lucent.com}
\author{B.S. Dennis$^{1}$}
\author{G.~Blumberg$^{1}$}
\email{girsh@bell-labs.com}
\affiliation{
$^{1}$Bell Laboratories, Lucent Technologies, Murray Hill, NJ 07974 \\
$^{2}$University of Illinois at Urbana-Champaign, IL 61801-3080}

\author{Seiki Komiya and Yoichi Ando}
\affiliation{Central Research Institute of Electric Power Industry, Komae,
Tokyo 201-8511, Japan
}

\date{\today}

\begin{abstract}
We study long wavelength magnetic excitations in lightly doped
\LSCO ($x \leq 0.03$) detwinned crystals.
The lowest energy magnetic anisotropy induced gap can be understood 
in terms of the antisymmetric spin interaction inside the 
antiferromagnetic (AF) phase.
The second magnetic resonace, analyzed in terms of in-plane
spin anisotropy, shows unconventional behavior within the AF state
and led to the discovery of collective spin excitations pertaining to
a field induced magnetically ordered state. 
This state persists in a 
9~T field to more than 100~K above the N\'{e}el temperature in
$x = 0.01$.
\end{abstract}


\maketitle

At short wavelengths the spin excitations in underdoped 2D cuprates are governed by the large antiferromagnetic (AF) superexchange, $J \approx 145$~meV, while in the long wavelength limit the magnetic energy scales are set by small anisotropy parameters~\cite{Kastner,KeimerPRB92,Coldea}.
In spite of the relative weakness, the impact of the low energy magnetism on the carrier and lattice dynamics in detwinned \LSCO crystals has been shown recently to be unexpectedly large.
Magnetic susceptibility data show a persistent spin anisotropy even outside the AF region~\cite{LavrovPRL01} while the transport properties revealed besides a sizeable anisotropy of the in-plane dc resistivity~\cite{AndoPRL02} also a large low temperature magnetoresistance~\cite{AndoPRL03}.
These effects call for an investigation of long wavelength magnetic excitations using a high resolution probe.
A magnetic field study is of particular interest because of the surprising discovery in x~=~0.01 \LSCO at room temperature of magnetic field induced structural changes~\cite{LavrovN}, phenomenon which makes this compound unique because of the existence of strong AF correlations.

Underdoped \LSCO crystals are slightly orthorhombic below 300~K and 
long range AF order exists for $x \leq 0.02$.
The layered structure allows for a XY spin anisotropy which in the
spin-wave approximation leads to an out-of-plane polarized gap,
$\Delta_{XY}$~\cite{PetersPRB88}.
The in-plane orthorhombicity and the spin-orbit coupling lead to an
antisymmetric Dzyaloshinskii-Moriya (DM) spin interaction in the CuO
planes which gaps the remaining Goldstone mode and leads to a
2$^{nd}$ in-plane polarized gap, $\Delta_{DM}$, at $k = 0$~\cite{PetersPRB88}.
The same spin-orbit interaction also allows the Raman coupling to
one-magnon excitations~\cite{Fleury}.
Each CuO plane has a weak perpendicular magnetic moment due to the DM
interaction but the interplane exchange orients the
canted moments antiferromagnetically as shown in Fig~1a.
An external magnetic field $\vec{H} \parallel \hat{c}$ leads to a
weak-ferromagnetic (WF) transition~\cite{ThioPRB88,AndoPRL03}
depicted in Fig.~1b.

In this Letter we study low energy Brillouin zone center magnetic
dynamics in lightly doped \LSCO as a function of doping, temperature
and magnetic field.
Two magnetic modes are observed in the AF phase.
The one at lower energies is the spin-wave gap induced by the
antisymmetric DM interaction and its anisotropic properties in
magnetic field can be well explained using a canonical form of the
spin Hamiltonian.
A new finding is a magnetic field induced mode (FIM) whose dynamics
allows us to discover a spin ordered state outside the AF order which
is shown to persist in a 9~T field as high as 100~K above the
N\'{e}el temperature T$_{N}$ in x~=~0.01.
We propose for the field induced magnetic order a state with a
net WF moment in the CuO plane and analyze the FIM in the context of
in-plane magnetic anisotropy.
\begin{figure}[b]
\epsfig{figure=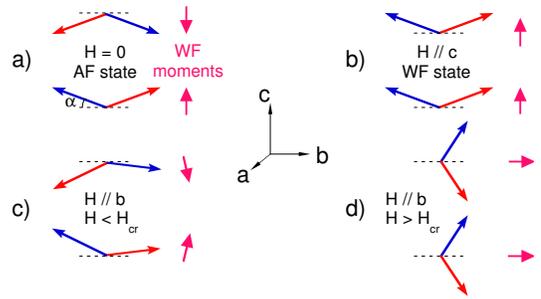,width=70mm}
\caption{
Cartoon showing spin orientations in two adjacent CuO planes.
(a) The 3D AF arrangement in zero applied field and the corresponding
orientation of the WF moments in each plane.
The canting angle $\alpha$ from the CuO planes is exaggerated.
(b) The WF state.
(c) Spin configuration for small fields $\vec{H} \parallel \hat{b}$.
(d) The proposed spin ordering obtained by further increasing the
field $\vec{H} \parallel \hat{b}$ in (c).}
\label{Fig.1}
\end{figure}

Detwinned single crystals of \LSCO with $x = 0 - 0.03$ were grown as
described in~\cite{LavrovPRL01}.
The N\'{e}el temperatures for the $x = 0$ and 0.01 crystals studied
here are 310 and 215~K respectively in zero field and a decrease of
T$_{N}$ on the average by 1~K/T (for x~=~0) and almost 4~K/T (for $x
= 0.01$) is observed for fields $\vec{H} \parallel \hat{b}$ axis.
We use the notation of the $Bmab$ structure for crystallographic axes.
We denote by $(e_{in} e_{out})$ polarization configurations, with
$e_{in/out}$ the direction of the incoming/outgoing photons.
$(RL)$ and $(RR)$ denote circular polarizations.
The crystals were mounted in a continuous flow optical cryostat and
the Raman data were taken with incident photon power of a few mW and 
$\lambda = 647.1$~nm focused onto $(ab)$ crystal surfaces.
The magnetic field data were taken with the cryostat inserted into
the horizontal bore of a superconducting solenoid.

Fig.~2 shows data from \LCO at 10~K for three directions of the
external magnetic field.
A sharp resonance is observed in zero field at 17~\cm-1.
This resonance disperses upwards (downwards) for $\vec{H} \parallel
\hat{a}$ ($\vec{H} \parallel \hat{b}$).
For $\vec{H} \parallel \hat{c}$-axis the mode disperses downwards
until $H_{WF} \approx 6$~T where the transition to the WF state
(Fig.~1b) takes place~\cite{ThioPRB88,AndoPRL03}.
In the 6~-~7~T range the resonance remains around 15~\cm-1 but
decreases in intensity with a concomitant appearance of another
feature around 21~\cm-1.
In Fig.~2e-f we show hysteretic loops for the 21 and 15~\cm-1 modes
intensities, very similar to the behavior of the (100) and (201)
magnetic Bragg peaks~\cite{KastnerPRB88}, reflecting the dynamics of
magnetic domains in the presence of small crystalline imperfections.

The nature of the mode shown in Fig.~2 can be understood by analyzing
the spin Hamiltonian~\cite{PetersPRB88}
\begin{eqnarray}
{\cal H} & = & \sum_{<i,j>} [ (J +\alpha) (S_{i}^{x} S_{j}^{x} +
S_{i}^{y} S_{j}^{y}) + J S_{i}^{z}
S_{j}^{z} + \vec{d} (\vec{S}_{i} \times \vec{S}_{j}) ] \nonumber \\ &
- & \vec{H} \sum_{i} \vec{S}_{i}
\label{DM}
\end{eqnarray}
where $J, \alpha, \vec{d}$, $\vec{H}$ and $\vec{S_{i}}$ are the
isotropic Heisenberg exchange, the in-plane anisotropy, the DM
vector, the external field and the spin on lattice site $i$.
We calculated the dispersion of the $k = 0$ spin-wave modes for
different magnetic field orientations at T~=~0~K by minimizing $\cal
H$ and linearizing the equations of motion~\cite{footnote1}.
The inset of Fig.~2d shows the results of this semi-classical
calculation for the behavior of the DM gap assuming a full moment on
Cu sites, $\vec{d} \parallel \hat{a}$ and $\Delta_{DM} = $~17~\cm-1.
The agreement between the experimental data and the calculation is
quantitative allowing us to assign this mode to $\Delta_{DM}$ and to
confirm the validity of the spin-wave approximation.
The similar field dispersion for $\vec{H} \parallel \hat{b}$ versus
$\vec{H} \parallel \hat{c}$ ($H < H_{WF}$) is intriguing because this
degeneracy does not follow from the model and it rather suggests
rotational symmetry with respect to the $a$-axis.
The energy of the DM gap is also in agreement with the DM gap value
inferred from inelastic neutron scattering (INS)
measurements~\cite{KeimerPRB92,KeimerZP}.
The softening of the DM gap for $\vec{H} \parallel \hat{b}$ explains
the decrease of T$_{N}$ and implies the possibility of a magnetic instability by further increasing the field in this configuration.
In principle our data allows also for a quantitative estimation of
the magnitude of other higher order spin
interactions~\cite{ShekhtmanPRL,Coldea} from magnetic gap
renormalization effects.
Using the expressions $\Delta_{DM} = 2.34 d$ and $J =
145$~meV~\cite{Kastner,Coldea} we obtain for \LCO the value $d = 0.92
\pm 0.013$~meV.
\begin{figure}[t]
\centerline{
\epsfig{figure=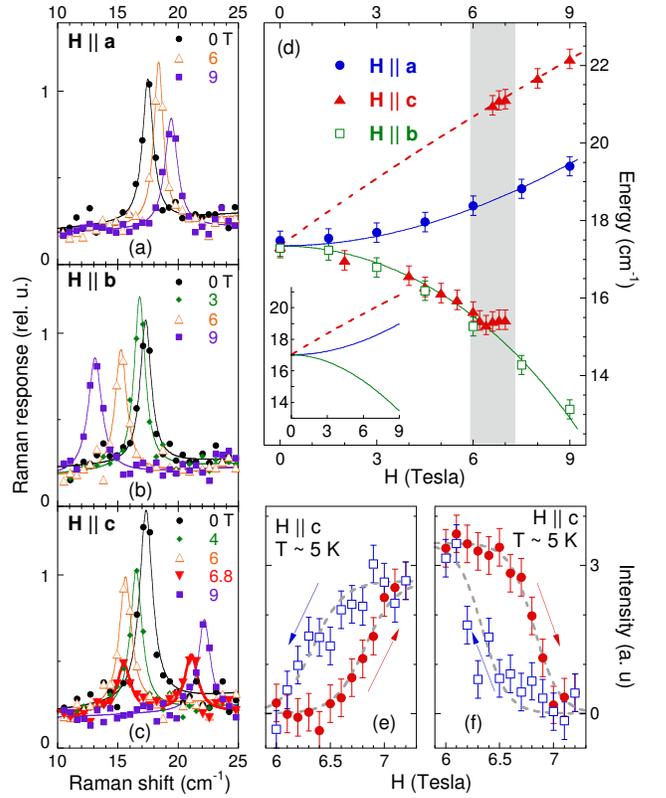,width=85mm}
}
\caption{
Panels (a-c) show 10 K $(RL)$ polarized Raman spectra of the DM gap 
in La$_{2}$CuO$_{4}$
with the external field $\vec{H}$ parallel to the $a$, $b$ and $c$-axes.
Solid lines are Lorentzian fits.
In (c) the 6.8~T spectrum shows the coexistence of the AF and the WF states.
(d) shows the field dependence of $\Delta_{DM}$ for $\vec{H}
\parallel \hat{a}$ (circles), $\vec{H} \parallel \hat{b}$ (squares)
and $\vec{H} \parallel \hat{c}$ (triangles).
The $\vec{H} \parallel \hat{c}$ data shows the transition to the WF
state depicted in Fig.~1b.
The continuous lines are fits with $\sqrt{\Delta_{DM}^{2} + \gamma
H^{2}}$ [$\Delta_{DM} = 17.35 \pm 0.25$~cm$^{-1}$, $\gamma_{H
\parallel a} = 0.96$ and $\gamma_{H \parallel b} = -1.65$~(cm
T)$^{-2}$] and the dashed line is a fit to the form
$\sqrt{\Delta_{DM}^{2} + \beta H}$ [$\beta = 22.6$~cm$^{-2}$T$^{-1}$].
Hysteretic loops of the lower (e) and higher (f) energy DM gaps at
the WF transition [shaded area in (d)].
Grey dashed lines in (e-f) are guides for the eye.
The inset in (d) shows the results of a semi-classical calculation of
the DM gap dispersions as described in the text, see also~\cite{footnote1}.
}
\label{Fig.2}
\end{figure}

Doping and temperature dependent properties of the DM gap are shown in Fig.~3.
Fig.~3a shows the gap as a function of doping at 10~K.
We observe that the DM gap is present only in the AF ordered region of the
phase diagram, being absent for $x \geq 0.02$.
The gap resonance found at 12.5~\cm-1 in x = 0.01 remains sharp but 
is weaker and has a strongly
renormalized energy compared to the undoped case.
As a function of temperature, the gap excitation
disappears below 
5~\cm-1 as we approach T$_{N}$ which
is indicative of a conventional magnetic soft-mode behavior (Fig.~3b-c).
The observed homogeneous renormalization with doping of
the DM gap energy at 10~K from x~=~0 to x~=~0.01 rules out a possible
macroscopic phase separation into $x \approx 0$ and $x \approx 0.02$
regions suggested in Ref.~\cite{MatsudaPRB02}.
Since the DM interaction is induced by lattice orthorhombicity, the
decrease of almost 30\% at T~=~10~K for the gap value between $x =
0$ and 0.01 is surprising compared to the much smaller decrease in
the orthorhombicity and relates it to strong sensitivity to hole
doping.
\begin{figure}[t]
\epsfig{figure=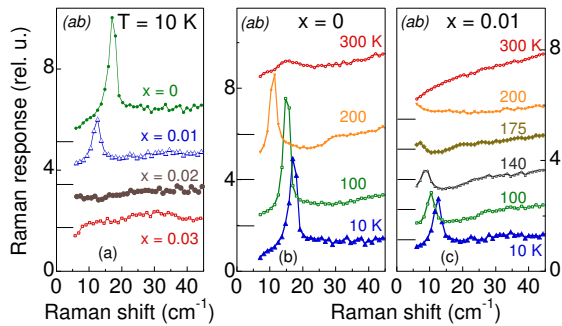,width=75mm}
\caption{
Doping and temperature dependence of the DM gap in \LSCO in $(ab)$
polarization for zero applied field.
(a) 10~K data for $x = 0 - 0.03$.
(b-c) Temperature dependent Raman spectra for $x = 0$ and $0.01$.
Data are vertically offset.}
\label{Fig.3}
\end{figure}
Our data suggest that the antisymmetric exchange interaction is
strongly competing with frustration and associated spin distortions
induced by hole doping~\cite{AharonyPRL88,Gooding}.

In Fig.~3b there is a broad peak at 300~K around
15~\cm-1 for x~=~0 which becomes a kink around 25~\cm-1 at 200~K and
disappears with further cooling.
The effects of magnetic fields on this excitation for \LCO are shown in Fig.~4.
At 10~K we observe for finite $\vec{H} \parallel \hat{b}$ a FIM around
40~\cm-1 which is sharp and hardens slightly up to 9~T.
At 230~K the FIM is broader and it softens with increasing field gaining spectral weight
from the lower energy side.
At 300~K we observe qualitatively similar behavior as at 230~K for $H
\leq 6$~T and beyond that value we see the emergence of two
independent peaks which harden further with field, see Fig.~4c.
Fig.~4b shows the total integrated intensity of the magnetic modes at
a given field.
The FIMs are not seen for any field direction other than $\vec{H}
\parallel \hat{b}$.

In \LCO the FIM dynamics marks two events.
The first seems to be a phase transition at 300~K and fields around 6~T.
This is indeed the case because we know that the N\'{e}el temperature in \LCO is around 310~K and
that the magnetic susceptibility $\chi_{b}$ shows T$_{N}$ decreasing at a rate of about 1~K/T.
Moreover, the narrow widths of the magnetic excitations above 6~T (2~\cm-1~$\approx$~0.25~meV) at temperatures more than two orders of magnitude higher (300~K~$\approx$~25~meV) argue strongly
for the collective nature of these excitations which correspond to another magnetically ordered state with a well defined gap in the excitation spectrum.
The second event, a crossover taking place between 230 and 10~K, is reflected in the opposite dispersion with field and different peak widths at these two temperatures.
\begin{figure}[t]
\epsfig{figure=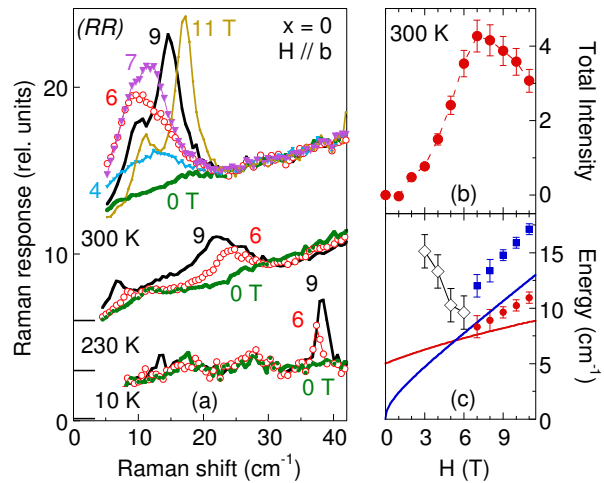,width=80mm}
\caption{
(a) Magnetic field ($\vec{H} \parallel \hat{b}$) dependence of $(RR)$
polarized Raman spectra in \LCO at several temperatures.
Data are vertically offset.
Total integrated intensities obtained by subtracting the 0~T spectrum 
in (a) from finite field data, panel (b), and the energies, panel 
(c), of the magnetic modes at 300~K.
The lines in (c) are calculated using Eq.~(1) with $\alpha = 0$, see text.}
\label{Fig.4}
\end{figure}
As for the doping dependence, except for a much weaker intensity (see
Fig.~5), we observed the same qualitative behavior in x~=~0.01
\LSCO.
The FIM is not seen (in fields up to 9~T) at any temperature for $x \geq 0.02$.

Fig.~5 shows temperature dependent $(RR)$ polarized spectra in a
9~T field $\vec{H} \parallel \hat{b}$ for x~=~0 and 0.01.
The crossover in \LCO takes place around 150~K, the temperature below
which the FIM width narrows.
Fig.~5c shows that the intensity of this excitation increases as we
approach T$_{N}$ and that around 300~K we observe the splitting due
to the occurrence of the field induced ordering.

The temperature dependence of the FIM across the N\'{e}el boundary
can be studied in the x~=~0.01 crystal which has a lower T$_{N}$.
At 9~T and below 180~K the behavior for x~=~0.01 is very similar to 
that in the undoped crystal showing a softening of the FIM as we warm 
to T$_{N}$.
The 200~K data show that the FIM has, similarly to the 295~K data 
for x~=~0, a low energy shoulder.
We believe that this weak excitation is the softened peak marked by
arrows at 250 and 275~K where the data shows the coexistence of two peaks.
The lowering in energy can be naturally explained in terms of soft mode
behavior above the magnetic transition while below T$_{N}$ this mode becomes the DM gap in Fig.~3c.
The two peaks above 180~K in x~=~0.01 \LSCO, which correspond in \LCO to the two features seen 
in Fig.~4c above 6~T, show that at 9~T the new magnetic order extends 
up to about 100~K above T$_{N}$.
In comparison, the strong feature at 200~K situated at 12~\cm-1
hardens only slightly with increasing temperature, see Fig.~5d.
\begin{figure}[t]
\centerline{
\epsfig{figure=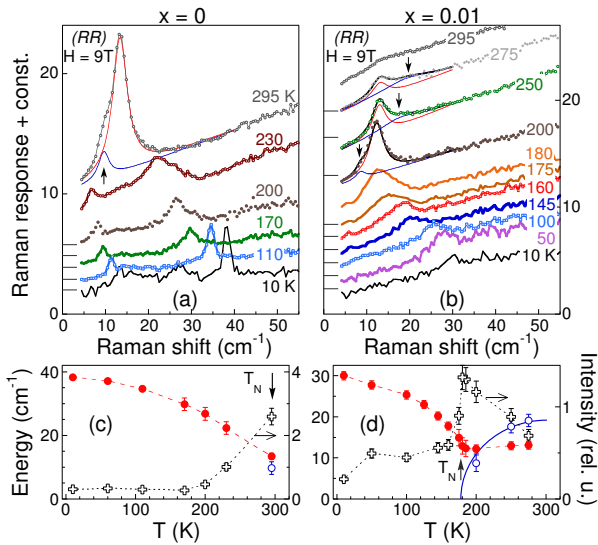,width=80mm}
}
\caption{
Temperature dependence of the field induced mode (FIM) in
La$_{2-x}$Sr$_{x}$CuO$_{4}$ for $x = 0$ (left) and $0.01$ (right).
(a-b) Data (vertically offset) in $(RR)$ polarization for $\vec{H}
\parallel \hat{b}$ at 9~T.
The continuous lines for T~=~295~K in (a), T~=~250 and 275~K in (b)
are two-Lorentzian fits to the data.
(c-d) Variation with temperature of the FIM energies (circles, left
scales) and intensities (crosses, right scales).
Empty circles correspond to the arrows in (a-b).
The blue line in (d) is a guide for the eye.
We also show by arrows the N\'{e}el temperatures for $\vec{H} \parallel
\hat{b}$ at 9~T in the two samples.}
\label{Fig.5}
\end{figure}
This panel also shows that the intensity of the FIMs is peaked at T$_{N}$ which is unexpected in a conventional picture where intensities of long wavelength gap modes scale with the AF order parameter as T$_{N}$ is approached from below~\cite{KeimerZP}.

A possible explanation for the FIM is its identification to the XY gap.
Support for this assignment is the presence of this mode only in x~=~0 and 0.01 \LSCO as well as the
comparison to INS data~\cite{KeimerPRB92,KeimerZP} which estimates $\Delta_{XY} \approx$~40~\cm-1 at 10~K.
Eq.~(1), which described well the DM gap, is also in support because it predicts (at T~=~0~K) a shift of about 6\% of the XY gap from 0 to 9~T, consistent with the small hardening we observe at 10~K in Fig.~4a.
As to the nature of the induced order, we propose a state like the
one depicted in Fig.~1d.
This is suggested by the magnetic susceptibility data which shows
that the moments on Cu sites remain confined in the $(bc)$
plane above T$_{N}$~\cite{LavrovPRL01} and also by recent magnetoresistance
measurements~\cite{AndoPRL03} which are consistent with a gradual
rotation of the WF moments.
A departure from a two-step transition \cite{ThioPRB90} with an $a$-axis 
spin-flop process occurring between the states shown in 
Fig.~1c-d is expected because the susceptibility $\chi_{a}$ is
the smallest below 300~K for x~=~0 and 0.01~\cite{LavrovPRL01} so the 
spins cannot partake of the field energy $- \chi_{a} H^{2} / 2$.

This scenario can also explain other observed features.
The crossover around 150~K shown in Figs.~4 and 5a may be
understood as a departure of the direction of the WF moments from
perpendicular to the $(ab)$ plane to a direction almost parallel to
the $b$-axis (see Fig.~1) where the XY anisotropy, weaker due to
temperature fluctuations, ceases to play a decisive role.
Physically,  this corresponds to the fact that the conventional
out-of-plane XY mode changes its nature as the WF moment rotates away
from the $c$-axis.
Prompted by this idea we calculated (solid lines in Fig.~4c) the
spin-wave dispersions using Eq.~(1) in the extreme case of $\alpha =
0$ and a small DM gap which still confines the moments in the
$(bc)$ plane.
Although finite temperature effects have to be taken into account, we
note that this simple estimation qualitatively reproduces the
experimental dispersions.
We also comment on the possible relevance of our findings to the
switch of orthorhombic axes in magnetic fields~\cite{LavrovN}.
If a state like Fig.~1d is realized (which we show in Fig.~5 to
persist to temperatures close to 300~K even for x~=~0.01) then the
magnetic force in an external field is significantly enhanced due to
the net in-plane magnetic moment.

The qualitative scenario we propose leaves as open questions the
finite Raman coupling to the FIM only for fields $\vec{H} \parallel
\hat{b}$ and also the surprising observation of its temperature dependent spectral weight being peaked at T$_{N}$.
However, if we assume that the FIM mode is an excitation other than
the XY gap, arising for instance as a result of the 4-sublattice
structure, then the common interpretation of the excitation around 40~\cm-1
found in several 2D layered AF's has to be reconsidered.

In summary, we discovered a field induced magnetically ordered
phase in detwinned x~=~0 and 0.01 \LSCO.
While the DM gap can be explained within the AF ordered state in the
framework of Eq.~(1), the behavior of the higher frequency field
induced modes both above and below T$_{N}$ requires an analysis
outside of the traditional interpretation of low energy magnetic
dynamics in \LSCO

We acknowledge useful discussions with A. N. Lavrov and M. V. Klein.


\begin{references}
\bibitem{footnote1}
Because Eq.~(1) neglects the interplane interaction it cannot be used
to account for the region $\vec{H} \parallel \vec{c}$ ($H < H_{WF}$).
\end{references}
\end{document}